\documentclass[aps,prl,twocolumn,superscriptaddress,showpacs,floatfix,letterpaper]{revtex4-1}%

\usepackage{graphicx}
\usepackage{bm}
\usepackage{latexsym}
\usepackage{epsf}
\usepackage{rotating}
\usepackage{epsfig,graphics,rotate,color}
\usepackage{wrapfig}
\usepackage{amssymb}
\usepackage{amsmath}
\usepackage{amsfonts}
\usepackage{subfigure}
\usepackage{array,hhline,dcolumn}%
\usepackage[normalem]{ulem}
\usepackage{color}
\usepackage{color}

\begin{document}

\title{Cross Section Measurements with Monoenergetic Muon Neutrinos}
\author{J. Spitz}
\affiliation{
Massachusetts Institute of Technology, Cambridge, Massachusetts 02139, USA
}

\begin{abstract}
The monoenergetic 236~MeV muon neutrino from charged kaon decay-at-rest ($K^+ \rightarrow \mu^+  \nu_\mu$) can be used to produce a novel set of cross section measurements. Applicable for short- and long-baseline accelerator-based neutrino oscillation experiments, among others, such measurements would provide a ``standard candle" for the energy reconstruction and interaction kinematics relevant for charged current neutrino events near this energy. This neutrino can also be exercised as a unique known-energy, purely weak interacting probe of the nucleus. A number of experiments are set to come online in the next few years that will be able to collect and characterize thousands of these events.     
\end{abstract}
\maketitle

\section{Introduction}
The muon neutrino charged current ($\nu_\mu$ CC) interaction is central to accelerator-based neutrino physics. However, despite the recent rapid progress in detection technology and analysis techniques, it is still quite difficult to measure the energy of $\nu_\mu$ CC events, especially in the case of an interaction with a nuclear target. Final state interactions (FSI), Fermi momentum, short-range correlations between nucleons, and the usually undetectable potential outgoing nuclear de-excitation gammas and neutrons can all work to convolute the reconstructed neutrino energy. This is before detector limitations, such as resolution, blindness to hadrons due to (e.g.) Cerenkov threshold, and event classification errors or ambiguity are even considered. Although a detector sensitive to the low-energy hadronic component of the final state can alleviate this at some level, many of these complications are simply inherent to the neutrino-nucleus system itself. As an example, a perfect detector's reconstruction of a 1~GeV $\nu_\mu$-nucleus CC quasi-elastic (CCQE, $\nu_\mu n \rightarrow \mu^- p$) event in terms of muon kinematics only provides an expected neutrino energy resolution of $\sim$20\%, with significant non-Gaussian asymmetric tails on either side of the true energy~\cite{martini_reco,martini_reco2,nieves_reco,mosel}. Although there are a number of predictions for the spectral smearing due to nuclear effects and the general direction of the convolution is understood, it is still highly non-trivial to correctly transform a reconstructed neutrino energy distribution into a true energy distribution with proper correlations and uncertainty estimates. Differences between neutrino and antineutrino events in terms of the nuclear physics that affects each can also be difficult to quantify.

A substantial amount of experimental effort has recently been directed toward measuring CC and neutral current (NC) neutrino cross sections at the hundreds~of MeV- and GeV-scale~\cite{PDG,formaggiozeller}. These measurements are interesting for the study of the neutrino-nucleus interaction itself, the neutrino as a probe of the nucleus (e.g. the strange spin component of the nucleon, $\Delta_s$~\cite{miniboonencel1,miniboonencel2}), and, perhaps most importantly, are essential for long-baseline neutrino oscillation programs and informing the simulations~\cite{nuwro,genie,neut} that such experiments rely on. The interaction cross section has also garnered a great deal of interest recently from theorists at the intersection of neutrino and nuclear physics, especially in the context of meson exchange currents and short-range correlations between nucleons~\cite{amaro,barbaro,budd,giusti,chanfray,nieves,sobcz}. 

The decay-at-rest of a positively charged kaon ($K^+ \rightarrow \mu^+  \nu_\mu$, BR=63.6\%~\cite{PDG}) produces a 236~MeV muon neutrino. A large sample of these monoenergetic neutrinos, interacting via the CC channel $\nu_\mu\mathrm{^{12}C} \rightarrow \mu^- X$, where $X$ is a proton and/or an excited nucleus, can be recorded to produce a standard candle for the $\nu_\mu$ and its interaction kinematics at and near this energy. As discussed in Ref.~\cite{spitz}, these charged kaon decay-at-rest neutrinos can also be used for a sensitive probe of high-$\Delta m^2$ oscillations indicative of a sterile neutrino. Unfortunately, these measurements are only really possible with neutrinos, rather than both neutrinos and antineutrinos, because of nuclear capture for negatively charged kaons. 

This article serves to point out the importance of measuring the differential and total cross sections associated with this unique, monoenergetic neutrino. For intense kaon decay-at-rest neutrino sources in general, a determination of the monoenergetic $\nu_\mu$ flux is complicated as the kaon production rate is difficult to accurately simulate and measure. Therefore, a precision, absolutely normalized cross section measurement is challenging. As discussed later, however, an exclusive interaction channel (for a carbon target, at least) can be used to determine the flux at the 10\% level in the case that enough events are collected. Either way, a single differential or double differential cross section measurement in terms of outgoing muon angle and/or momentum, with detailed shape information, is valuable. Given a known neutrino energy, a precision differential cross section measurement in energy transfer, as is common for electron scattering experiments, is also available. No such measurement exists for $\nu_\mu$ CC interactions. After discussing the relevance of these measurements in a number of physics applications, a set of potential experimental locations and detection technologies are considered.

\section{Physics with neutrinos from kaon decay-at-rest}
Along with using the known-energy neutrinos as a unique probe of the nucleus, and as a test of our theoretical description of neutrino-nucleus interactions at this energy, a set of 236~MeV $\nu_\mu$ CC cross section measurements can be utilized in a number of ways. 

Long-baseline neutrino oscillation experiments employ near and far detectors for measuring the $L/E$-dependent mixing probability of a beam of originally $\nu_\mu$ or $\overline{\nu}_\mu$. These experiments are able to run in either neutrino or antineutrino mode by changing the polarization of their beamline magnets to focus $\pi^+$ ($\rightarrow \mu^+  \nu_\mu$) or $\pi^-$ ($\rightarrow \mu^-  \overline{\nu}_\mu$). The near detector, typically hundreds of meters from the source, provides a description of the initial, pre-oscillation composition of the mostly pure $\nu_\mu$ or $\overline{\nu}_\mu$ beam. The far detector, typically hundreds of kilometers from the source, probes the beam for $\nu_\mu$ and $\overline{\nu}_\mu$ disappearance as well as $\nu_e$ and/or $\nu_\tau$ appearance. These experiments are sensitive to the $\theta_{23}$ octant, the orientation of the mass hierarchy, and the neutrino \textit{CP}-violating phase $\delta_{\mathrm{CP}}$, among other physics.

The $\nu_\mu$ or $\overline{\nu}_\mu$ CCQE interaction is often utilized both as the signal channel for the disappearance measurements and to constrain predictions relevant for the appearance channels. Given the $L/E$ oscillation dependence, reconstructing the energy of these neutrinos is obviously vital to extracting the mixing parameters. Further, an appropriate comparison between the near and far detector event rate as a function of reconstructed neutrino energy (and/or lepton kinematics, as in Ref.~\cite{t2k_nueapp}) requires knowledge of the interaction cross section because the energy composition of the beam is different at each site, even in the case that both detectors are on-axis. Along with near-far flux differences due to oscillations, this is due to the fact that the near detector is exposed to a range of incident neutrino angles (and corresponding kinematics) from pions decaying in flight at different locations in the decay pipe, while the far detector is effectively exposed to a point source of neutrinos. The possibility of systematic differences between the near and far sites in composition, size, detection technology, etc. can lead to further reliance on knowledge of the underlying cross section as well. 

As an example, the uncertainties associated with the neutrino interaction dominate the systematics on the predicted number of signal $\nu_e$ events in the T2K long-baseline neutrino oscillation experiment~\cite{t2knim,t2k_nueapp}. These uncertainties, especially those related to the $\nu_\mu$/$\nu_e$ ($\overline{\nu}_\mu$/$\overline{\nu}_e$) cross section ratio~\cite{huberpaper}, will likely be among the leading sources of systematic uncertainties in future $\delta_{\mathrm{CP}}$ programs, and there are multiple experiments around the world with the goal of improving our knowledge of the kinematics of the outgoing charged lepton and the neutrino/antineutrino cross section on various nuclear targets in the long-baseline energy regime. The situation is such that it may be advantageous for future long-baseline experiments to tune their characteristic neutrino fluxes to better align with regions of low cross section and/or energy reconstruction uncertainty. 

The few-hundreds-of-MeV neutrino energy range is an interesting possible place of emphasis for future long-baseline experiments for a number of reasons.  Along with tuning closer to the second or third oscillation maxima, depending on baseline, in an attempt to better distinguish between different $\delta_{\mathrm{CP}}$ scenarios and perhaps moving toward a more well-understood cross section and neutrino energy reconstruction region, given a precise set of (e.g.) 236~MeV-based measurements, emphasizing a lower beam energy can also prove advantageous because of the smaller resonant background and reduction of complications due to short-range correlations. Short-range correlations are not expected to play a large role at these relatively low neutrino energies~\cite{martini_reco2,nieves_reco}. However, in the case of a statistics-limited measurement, as compared to a background- or interaction-systematics-dominated one, a lower energy may not be optimal due to the reduction in cross section, and therefore event rate, as well as the lack of muon reconstruction abilities below 54~MeV for water-based Cerenkov detectors such as SuperK/HyperK. Further, it is not clear if the benefit of the cross section measurements outlined here, in a move to lower energy, can outweigh the comparatively poor theoretical understanding of the $\nu_\mu$/$\nu_e$ cross section ratio in this low energy part of the QE regime.

Beyond the general comments above, the impact of a set of 236~MeV cross section measurements is potentially wide ranging, especially in the case of a future experiment that relies on cross section knowledge in the few-hundred-MeV neutrino energy region. As an example, the European Spallation Source Neutrino Super Beam (ESS$\nu$SB) long-baseline neutrino oscillation project aims to produce a 5~MW, 2~GeV proton beam and combine it with a water Cerenkov based detector 300-600~km away in order to measure $\delta_{\mathrm{CP}}$~\cite{ESSnu}. In neutrino mode, the $\nu_\mu$ flux peaks at about 225~MeV. There is also an idea to employ simultaneous, high power 8- and 60-GeV proton beams with a 200~kt water Cerenkov detector to obtain sensitivity to ``low" energy (0.2-1.5~GeV) $\nu_\mu \rightarrow \nu_e$ oscillations at the second oscillation maximum at a distance of 1300~km. Such an experiment would provide a precise determination of $\delta_{\mathrm{CP}}$ that is largely independent of the mass hierarchy orientation~\cite{bishai}. A cross section measurement with kaon-induced neutrinos is also quite relevant for a future $\beta$-beam with a $\gamma$ near 100 and/or a 3.5~GeV Super Proton Linac (SPL) super-beam from CERN; both sources will produce neutrino fluxes in the few-hundred-MeV range~\cite{beta_spl}. The $\beta$-beam-SPL combination has been identified as a future option that will have greater sensitivity to $\delta_{\mathrm{CP}}$ than any other super-beam or $\beta$-beam concept, second only to a neutrino factory~\cite{coloma}. 

As is true for all neutrinos in the long-baseline energy regime, nuclear effects play a large role for 236~MeV $\nu_\mu$ interactions and, despite the value in the cross section measurements outlined, it should be stated that this neutrino energy is a challenging one to deal with theoretically. The neutrino energy, or rather the characteristic energy transfer, is right at the transition between our neutrino-on-nucleus and neutrino-on-nucleon frameworks. The impulse approximation, in which it is assumed that the neutrino interacts with a single nucleon, breaks down at these lower energies and the distinction between a pre-FSI pure CCQE interaction ($\nu_\mu n \rightarrow \mu^-  p$) and an ``absorption" interaction (on carbon, for example, $\nu_\mu  \mathrm{^{12}C} \rightarrow \mu^- X$) becomes blurred. Indeed, measurements of the monoenergetic neutrino may shed light on this important transition and inform the theoretical representations. 

\section{Experimental locations}
There are a number of proton fixed target experimental locations around the world where these cross section measurements are possible. The main requirement, other than a capable existing or planned detector, is that the primary proton energy exceed $\sim$3~GeV for adequate kaon production. The 1.4~MW, 1~GeV Spallation Neutron Source at Oak Ridge and the future 5~MW, 2~GeV European Spallation Source are therefore not considered here. In addition, given that a decay-at-rest source produces an isotropic flux of neutrinos, the detector needs to be reasonably close ($\lesssim$100 m) to the source. The NuMI beam dump at Fermilab and the JPARC Materials and Life Science Facility (MLF) spallation source, in combination with a set of nearby planned detectors, are considered as possible experimental locations here. 

The 120~GeV NuMI beamline terminates for all non-neutrinos at a beam dump, 720~m downstream of the target~\cite{numi}. Given the target's two interaction lengths, as much as 14\% of the beam is passed on to this beam stop. The NuMI beam dump therefore provides a significant source of kaon decay-at-rest neutrinos and there are a set of nearby detectors that are sensitive to the $\nu_\mu$ CC interaction at 236~MeV: MiniBooNE, running since 2002 and located about 85~m from the NuMI dump, is a Cerenkov- and scintillation-based mineral oil detector~\cite{mini_nim}, and MicroBooNE, running in 2014 and located about 102~m from the dump, is a Liquid Argon Time Projection Chamber (LArTPC)~\cite{uboone_prop}. While MiniBooNE has probably collected thousands of these from-NuMI events~\cite{minibooneatnumi}, it is likely difficult to unambiguously identify them as monoenergetic ones given the detector's propensity for lepton-only reconstruction for this class of interactions. Further, the pion decay-in-flight $\nu_\mu$ ``background", largely coming from the NuMI decay pipe, likely makes the bump difficult to pick out. However, MiniBooNE$+$, an experimental proposal to add liquid scintillator to the existing MiniBooNE detector~\cite{minibooneplus}, may be able to enhance the energy reconstruction abilities enough to resolve the from-kaon peak. With full kinematic reconstruction abilities, as with the MicroBooNE LArTPC, the decay-in-flight background can be reduced significantly with reconstructed neutrino energy and direction requirements; this experiment is an attractive future location for these measurements. 

It is worth noting that the SciBooNE experiment has also likely collected a significant number of these monoenergetic neutrinos, originating at the Booster Neutrino Beamline (BNB) dump, but the decay-in-flight $\nu_\mu$ flux from the BNB decay pipe is more than an order of magnitude higher in the relevant energy region~\cite{scibooneccinc}. This will also be true for the future LAr1-ND experiment~\cite{lar1nd}, proposed to be located at the SciBooNE hall in the BNB. 

The JPARC-MLF is host to another intense source of kaon decay-at-rest neutrinos coming from 3~GeV protons on a mercury target. The eventually 1~MW source features only a small decay-in-flight background component as it is nominally used for spallation neutron production rather than as a conventional neutrino beamline. The mercury target is basically surrounded on all sides by concrete and iron, and the large majority of pions, muons, and kaons created quickly come to rest and capture or decay. Currently, there are plans to place a 50~ton fiducial volume gadolinium-loaded liquid scintillator (LS) neutrino detector 17~m from the source for a sterile neutrino search there~\cite{harada}. Such a detector can also be used to perform the measurements described here. 

\section{Detection}
We survey the detection technologies associated with the JPARC-MLF (LS) and MicroBooNE (LArTPC) experiments when considering these cross section measurements. Although a Cerenkov-based detector with muon-only reconstruction proficiency could potentially pick out the monoenergetic bump in reconstructed energy due to the 236~MeV neutrino, especially in the absence of a significant non-monoenergetic background, both technologies considered here have better neutrino energy reconstruction capabilities, mainly because of their ability to reconstruct the low-energy nucleonic component of these events; LS and LArTPC technology are simply more suitable for making sure that the events being evaluated are indeed coming from charged kaon decay-at-rest. Of course, what is learned from these measurements as well as their applicability toward future oscillation programs is highly dependent on which nuclear target is chosen.

The detection of $\nu_\mu$ CC events up to $\sim$260~MeV with LS in the Liquid Scintillator Neutrino Detector (LSND) experiment is discussed at length in Ref.~\cite{lsnd_numu_2002}. A combination of both scintillation and Cerenkov light signals can provide directional, calorimetric, and particle identification information for reconstructing the events. The muon is identified by requiring a delayed coincidence with a characteristic decay electron and can also be distinguished with a Cerenkov signal since nearly 90\% of monoenergetic events produce a muon above the 36~MeV kinetic energy threshold [in the commonly-used linear alkyl-benzene based LS]. Stopping $\mu^-$ are captured 8\% of the time on carbon in the LSND detector~\cite{lsnd_numu_2002}. A veto, in combination with beam timing, can render the steady state background, mainly coming from cosmic ray muon decay in the detector, negligible. Notably, the JPARC-MLF source features an extremely tight beam window with two 80~ns wide pulses of protons 540~ns apart at 25~Hz, resulting in a steady state rejection factor of $4\times10^{-6}$. It is also expected that the JPARC-MLF LS detector will feature faster electronics than LSND, although it is difficult to estimate the achievable muon momentum and angular resolutions until the detector parameters, such as photo-coverage and time resolution, are finalized. For reference, LSND's 25\% photo-coverage resulted in a muon directional reconstruction resolution of about $12^\circ$ for muons above threshold and an energy resolution of better than 10\% at $T_{\mu}=100~\mathrm{MeV}$~\cite{louis_communication,lsnd_numu_2002}. 

In a best case scenario, the contributions of the scintillation- and (usually) Cerenkov-ring-producing muon and scintillation-only proton (or protons, since FSI and correlations can result in multiple ejected nucleons), can be separated in LS for a more precise measurement of the differential cross sections, especially in terms of reconstructing the momentum of the muon. In practice, however, this is difficult and will likely require successfully modeling the light production of both the outgoing proton(s) and nuclear de-excitation gammas.   

In a LArTPC, the charged particles created in a neutrino interaction, the reconstruction of which is required in order to infer the energy and flavor of the neutrino itself, propagate through the liquid argon medium and create trails of ionization along their paths. An electric field is imposed in the liquid argon volume and the trails are drifted through the noble liquid toward a set of sensing electrodes. The signals in time captured by the electrodes, usually in the form of a set of wire planes oriented at an angle with respect to one another, provide a complete three-dimensional image of the neutrino event. Calorimetric information is available as the ionization collected by the electrodes is related to the amount of energy deposited along the charged particle tracks. Scintillation light (128~nm) is also produced readily as the charged particles ionize atoms; argon's high scintillation yield is useful for detecting this aspect of the interaction as well, although a wavelength shifter is required in conjunction with photomultiplier tubes to shift the light into the visible spectrum and detect it.  With sensitivity to de-excitation gammas, neutrons, protons down to the few-tens-of-MeV level, and precise calorimetric reconstruction abilities, LArTPC technology is attractive for detecting and characterizing 236~MeV $\nu_\mu$ CC events.

\begin{figure}[h]
\begin{centering}
\includegraphics[height=2.4in]{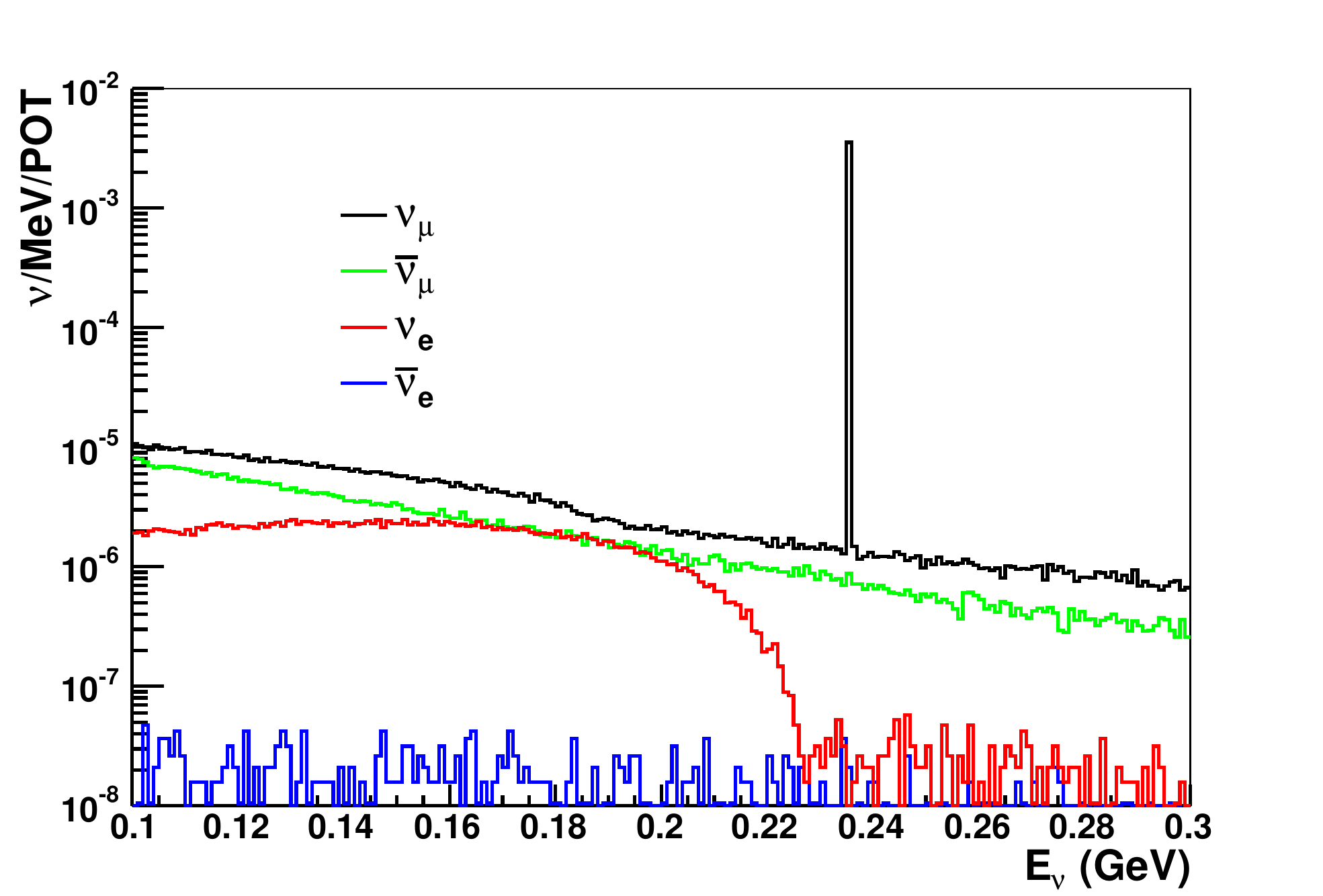} 
\caption{The neutrino flux from 100-300~MeV provided by the 3~GeV proton-on-mercury JPARC-MLF source. The 236~MeV charged kaon decay-at-rest daughter $\nu_\mu$ is easily seen.}
\label{mlf_flux}
\end{centering}
\end{figure}

Table~\ref{eventrate} shows the expected number of monoenergetic $\nu_\mu$ CC events in both MicroBooNE and the LS detector at the JPARC-MLF. The MicroBooNE event rate estimate assumes two years of running NuMI in neutrino mode at 700~kW ($6\times 10^{20}$~POT/year), consistent with the Fermilab roadmap. Interestingly, NuMI neutrino mode and antineutrino mode each provide a similar flux of monoenergetic neutrinos. The JPARC-MLF event rate estimate assumes four years of running with a 1~MW beam and 4000~hours/year of operation, or $3\times 10^{22}$~POT/year, consistent with Ref.~\cite{harada}. The neutrino flux at each location has been determined using GEANT4~\cite{geant4} (and FLUKA~\cite{fluka} also, in the case of NuMI) simulations of the sources, noting that kaon production is highly uncertain at both locations. As an example, the kaon-induced monoenergetic $\nu_\mu$ production at the 3~GeV JPARC-MLF source is 0.0035~$\nu_\mu /\mathrm{proton}$ with GEANT4 but is found to be about 75\% higher with the LAQGSM/MARS (MARS15) software package~\cite{mars}. The GEANT4 results are used here in order to be conservative. The event rate estimates also assume a $\nu_\mu$ CC cross section of $1.3\times10^{-39}~\mathrm{cm}^2/\mathrm{neutron}$, consistent with the NuWro neutrino event generator for interactions on both carbon and argon at 236~MeV~\cite{nuwro} and the theoretical predictions~\cite{volpe1}. The expected neutrino flux from the JPARC-MLF source in all directions, without regarding potential detector location, in the energy range 100-300~MeV can be seen in Fig.~\ref{mlf_flux}. The 236~MeV $\nu_\mu$ and three-body kaon decay ``$K_{e3}^+$" ($K^{+} \rightarrow \pi^0  e^+  \nu_{e}$, BR=5.1\%) $\nu_e$ distributions are obviously quite prominent. 

\begin{table*}[t]
      \begin{tabular}{|c|c|c|c|c|} \hline \hline
Detector (source)& Target (mass) & Exposure & Distance from source  & 236~MeV $\nu_\mu$ CC events  \\  \hline 
MicroBooNE (NuMI dump)
 & LAr (90~ton) & $1.2\times 10^{21}$ POT (2 years) & 102 m & 2300  \\
Liq. scint. (JPARC-MLF) & Gd-LS (50~ton) & $1.2\times 10^{23}$ POT (4 years) & 17 m & 194000 \\ \hline \hline
\end{tabular} 
\caption{The expected monoenergetic $\nu_\mu$ CC event rate at two experimental locations along with the beam exposure and detector assumptions. }
\label{eventrate}
\end{table*}

\begin{figure}[h]
\begin{centering}
\includegraphics[height=2.4in]{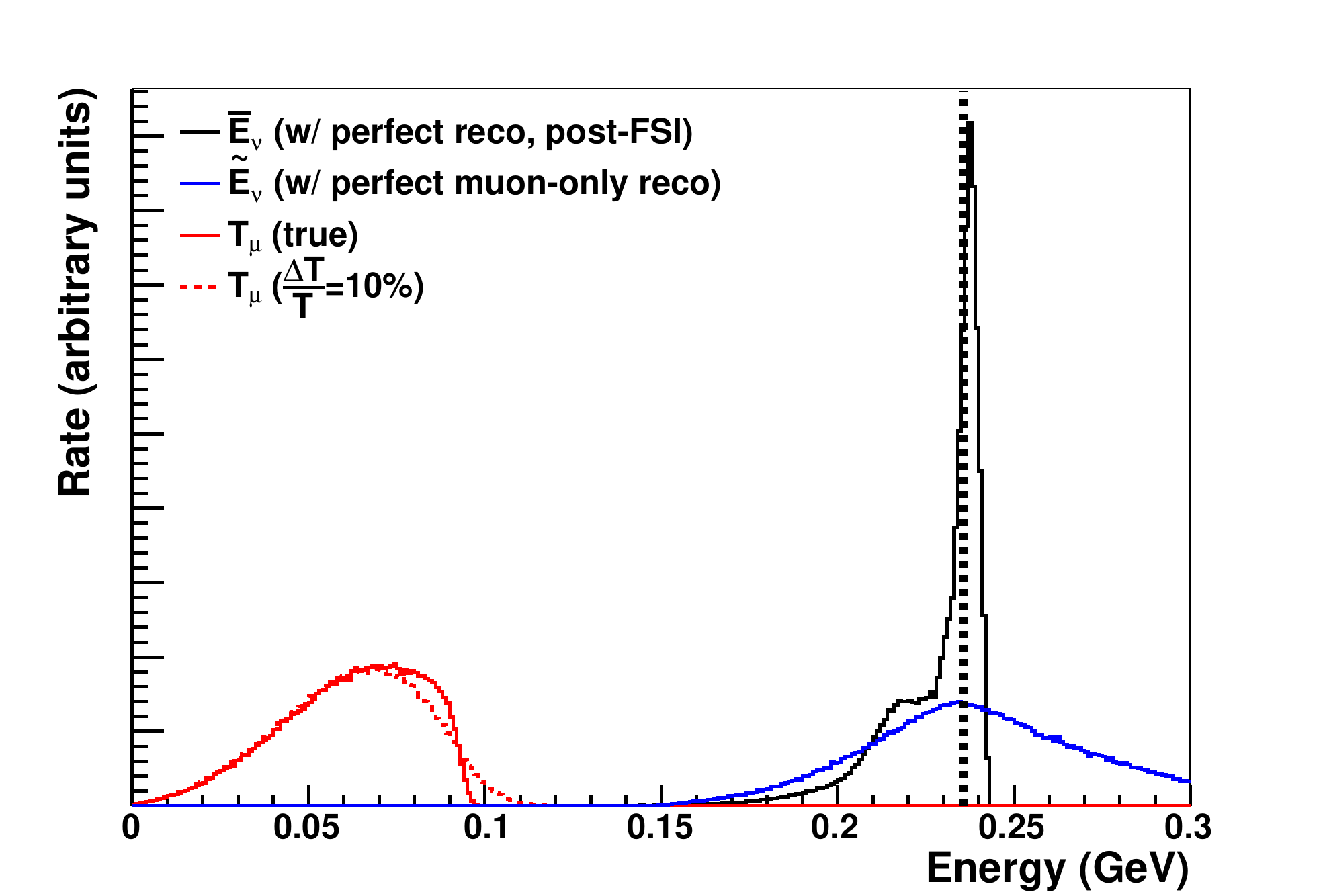} 
\caption{The NuWro simulation results of 236~MeV $\nu_\mu$ (from $K^{+} \rightarrow \mu^+ \nu_\mu$) CC interactions on carbon. The kinetic energy of the outgoing muon, given a set of detector resolution assumptions, is shown. The reconstructed neutrino energy available with a perfect detector is also visible, noting that de-excitation gammas and neutrons are considered missing energy here. The reconstructed neutrino energy with perfect muon-only tracking is also shown for reference.}
\label{muon}
\end{centering}
\end{figure}

\begin{figure}[h]
\begin{centering}
\includegraphics[height=2.3in]{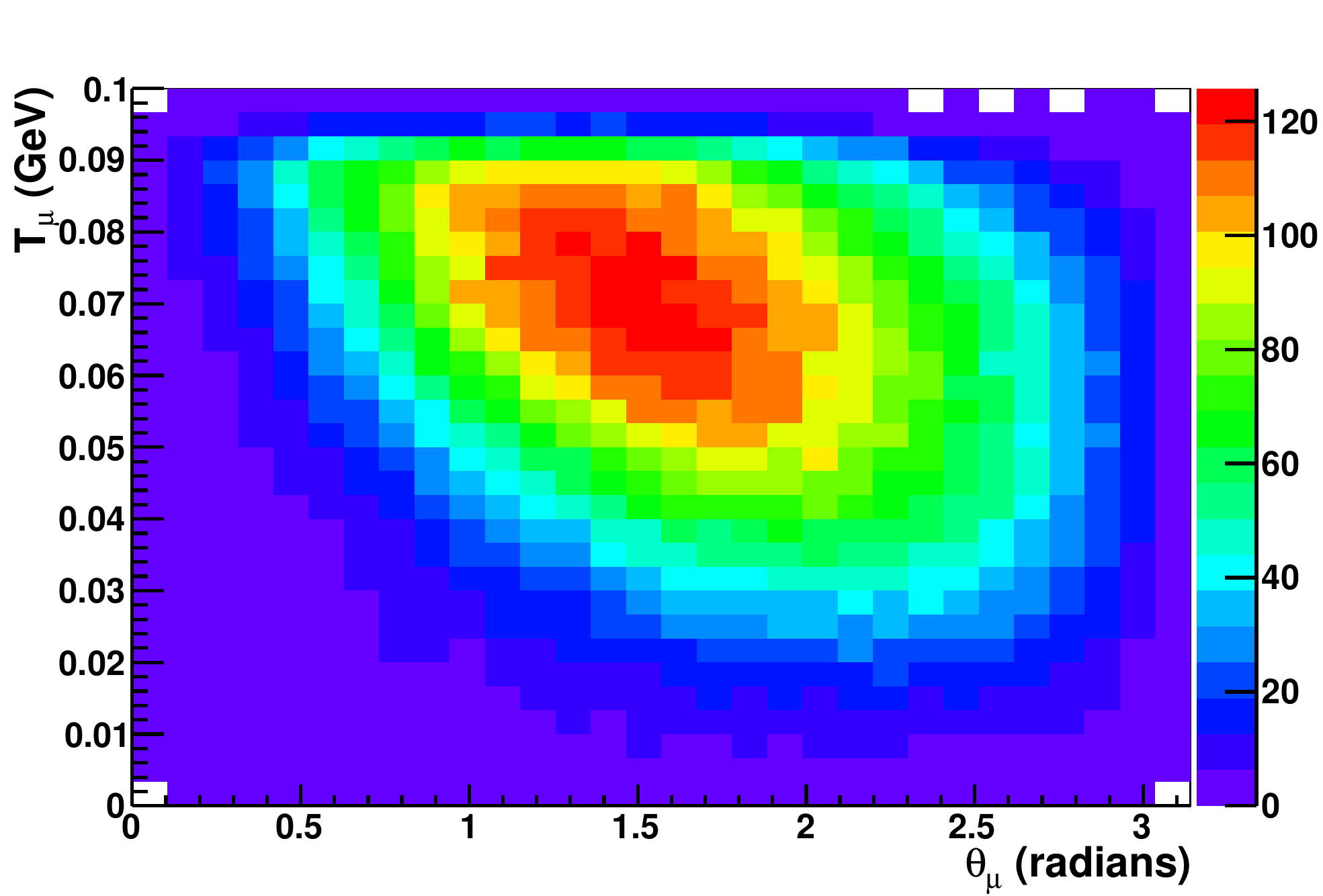} 
\caption{The kinematics of the muon from 236~MeV $\nu_\mu$ CC interactions on carbon according to NuWro. The z-axis units are arbitrary.}
\label{muon_kin}
\end{centering}
\end{figure}

The NuWro neutrino event generator has been used here in order to simulate 236~MeV $\nu_\mu$ CC interactions on carbon and argon. The simulation provides an idea of what can be expected from these neutrinos, although the employed impulse approximation is known to simulate neutrinos, especially $\nu_\mu$, poorly at these relatively low energies. NuWro is used because it contains a spectral-function-based simulation for both nuclei~\cite{nuwro_c}. The results of the neutrino-on-carbon simulation are shown in Fig.~\ref{muon}. The kinetic energy of the muon is seen along with a Gaussian smeared energy, given an arbitrary 10\% detection resolution. Also, the post-FSI reconstructed neutrino energy $\overline{E}_\nu$ ($=$$E_\mu + \sum_i^n T_{i,\mathrm{proton}} + S_p$, where $n$ is the number of protons and $S_p=16$~MeV is the proton separation energy for $^{12}$C) with a perfect detector, after considering neutron and de-excitation gammas non-reconstructable, is shown. The separation energy for a single proton only is used for simplicity. The apparent bimodal shape of the distribution is due to the shell structure of the nucleus and the energy levels of the neutron within the spectral function implementation. The reconstructed energy with a perfect muon-only detector $\tilde{E}_\nu$, given the usual two-body kinematics CCQE formula, assuming target nucleon at rest and a binding energy of 34~MeV, is also shown for reference. The shape of the expected 236~MeV $\nu_\mu$ CC event rate distribution in muon angle and kinetic energy, as simulated with NuWro, is shown in Fig.~\ref{muon_kin}.

The main requirement for a valuable monoenergetic $\nu_\mu$ cross section measurement is that the signal interaction is properly identified as such. This determination relies on the ability to precisely reconstruct the energy of the neutrino, with the actual energy resolution needed depending on the background non-monoenergetic $\nu_\mu$ flux in the energy region of interest. In the case of the JPARC-MLF source, for example, the monoenergetic cross-section-weighted flux, without regard for detector location, will be a factor of $\sim$30 times higher than the integrated background in an arbitrary true energy window of 80~MeV around 236~MeV. The actual signal-to-background at the detector location will likely be significantly higher than this, given the tentative backward orientation of the detector relative to the primary proton beam direction, according to Ref.~\cite{harada}, and resulting decrease in the decay-in-flight component at the detector. For example, the ratio increases to $\sim$180 in the case that only neutrinos with $\cos \theta_z < 0$ (where $+z$ is the primary proton direction) are considered. Regardless, if a $\nu_\mu$ CC event is identified from this source, with even modest energy resolution, one can be fairly confident that it is monoenergetic. 

The favorable signal-to-background ratio at the JPARC-MLF may afford the ability to perform these cross section measurements with a water Cerenkov detector, capable of providing lepton-only kinematic reconstruction, for a cross section measurement directly applicable to the Super-K and Hyper-K detectors~\cite{HyperK_LOI} within the T2K long-baseline program, and water-based devices in general. However, the 236~MeV $\nu_\mu$ is at the low end of neutrino energies relevant for T2K ($E_\nu^{\mathrm{peak}}\sim600~\mathrm{MeV}$). Further, the water Cerenkov threshold for muons is 54~MeV in kinetic energy so one-third of the signal is nominally irretrievable (see Fig.~\ref{muon}) and, when considering muons in the T2K $\nu_\mu$ disappearance analysis, Super-K's threshold (200~MeV/c, due to particle identification requirements at low energy) is above the characteristic momenta for monoenergetic events~\cite{t2k_muon_disapp}.

The large JPARC-MLF signal-to-background ratio in an arbitrary true neutrino energy window of 80~MeV around 236~MeV decreases to $\sim$1 in the case of the NuMI beam dump source given the substantial pion decay-in-flight background $\nu_\mu$ flux. Along with precisely reconstructing the neutrino energy in order to reduce background, this issue can be mitigated somewhat with the ability to reconstruct the direction of the incoming neutrino since the large majority of monoenergetic $\nu_\mu$ will be coming directly from the dump rather than the NuMI decay pipe or target station. This requires the ability to reconstruct the low-energy hadronic component of the interaction at a reasonable level as the outgoing muon and incoming neutrino direction are rather poorly correlated. Notably, the ArgoNeuT LArTPC detector has demonstrated the ability to reconstruct protons down to 21~MeV kinetic energy~\cite{tingjun}. While it is currently not clear exactly how well MicroBooNE will be able to reconstruct the direction of 236~MeV events, especially given that nuclear effects can distort the direction of the outgoing nucleon(s), the detector's orientation relative to the NuMI beamline and dump, with the NuMI target and dump separated by about 110$^\circ$, will make it easier to distinguish between neutrinos from the different sources. Even if this is not achievable, the  decay-in-flight background contribution to a monoenergetic $\nu_\mu$ measurement can be constrained by considering events outside of the relevant energy region. Data collection with a proton beam plug or lengthened target, which would substantially reduce the monoenergetic flux at MicroBooNE, could also allow for an $in-situ$ measurement of the background component. 
  
\section{Other opportunities with kaon decay-at-rest neutrinos}
Aside from measurements of the $\nu_\mu$ CC differential and total cross sections described above, from-kaon neutrinos provide a number of other physics opportunities as well. Two of these are described below: (1) A measurement of an exclusive channel can deliver a precise determination of the neutrino flux at 236~MeV, and (2) the charged kaon decay $K_{e3}^+$ can provide a significant sample of $\nu_e$ events in an energy range relevant for accelerator-based oscillation measurements. 

Despite the complications discussed above with regard to the transition between neutrino-on-nucleon and neutrino-on-nucleus scattering, the contribution from the absorption part of the inclusive interaction can be seen as something of a windfall, at least when considering a carbon target, given the significant $\nu_\mu  \mathrm{^{12}C} \rightarrow \mu^-  \mathrm{^{12} N_{\mathrm{gs}}}$ exclusive cross section at this energy. This exclusive reaction, which can be purely identified with a triple coincidence of the muon, the decay daughter electron, and the positron from the $\beta$ decay of  $\mathrm{^{12} N_{\mathrm{gs}}}$, has a well-predicted cross section at 236~MeV of $\approx$$ 7\times10^{-41}~\mathrm{cm}^2/\mathrm{nucleus}$~\cite{engel} which results in an event rate of about 1\% relative to the inclusive channel. Although this rate is comparatively low, it still would provide nearly 2000~monoenergetic events in four years of running with the JPARC-MLF detector. While the theoretical cross section prediction for the inclusive channel is highly uncertain, with models differing by up to $\sim$$25\%$ for LSND's flux-averaged cross section ($<E_\nu>=156~\mathrm{MeV}$)~\cite{lsnd_numu_2002,kolbe,volpe1,hayes}, the exclusive channel is well known, with differences between the various shell-model-based predictions at the level of only 10\% at 250~MeV~\cite{lsnd_numu_2002,engel}. The exclusive prediction is more precise because it relies on form factors arrived at with measured values of the related electroweak transition probabilities ($\beta$ decay and muon capture)~\cite{volpe1}. Given the reliable cross section prediction, this exclusive channel can be used for the absolute flux determination at this energy and therefore for reporting precisely normalized differential and total cross section measurements. 

Along with the monoenergetic $\nu_\mu$, kaons yield another potentially important source of neutrinos as well. While $\nu_\mu$ CC cross section measurements in the energy range 53-500~MeV are quite sparse~\cite{lsnd_numu_2002,ANL,minibooneccqe2,scibooneccinc}, $\nu_e$ cross section measurements above 53~MeV are non-existent. A significant number of $\nu_e$ CC events can be collected from the JPARC-MLF source (via $K_{e3}^+$), noting that the $\nu_e$ cross section is about 25\% higher than $\nu_\mu$ at these energies and the $\nu_e$ cross section has a significantly lower energy threshold due to the muon-electron mass difference. Although the $\nu_e$ are not monoenergetic, the flux does cut off sharply at $\approx$225~MeV and has a characteristic energy shape. With regard to applying such a measurement to T2K, the neutrino energy is low but well within relevance, given that T2K requires the reconstructed electron momentum to exceed 100~MeV/c in their appearance analysis~\cite{t2k_nueapp}. Indeed, one of T2K's $\nu_e$ appearance candidates has a reconstructed energy of $\approx$150~MeV~\cite{t2k_nueapp}. Since there are no $\nu_e$ CC measurements at these energies and this channel, along with its antineutrino analog, represents the actual signal for a long-baseline neutrino oscillations experiment's $\delta_{\mathrm{CP}}$ measurement, as well as numerous short-baseline electron-flavor appearance searches, this sample might prove quite useful. 

\begin{figure}[h]
\begin{centering}
\includegraphics[height=2.4in]{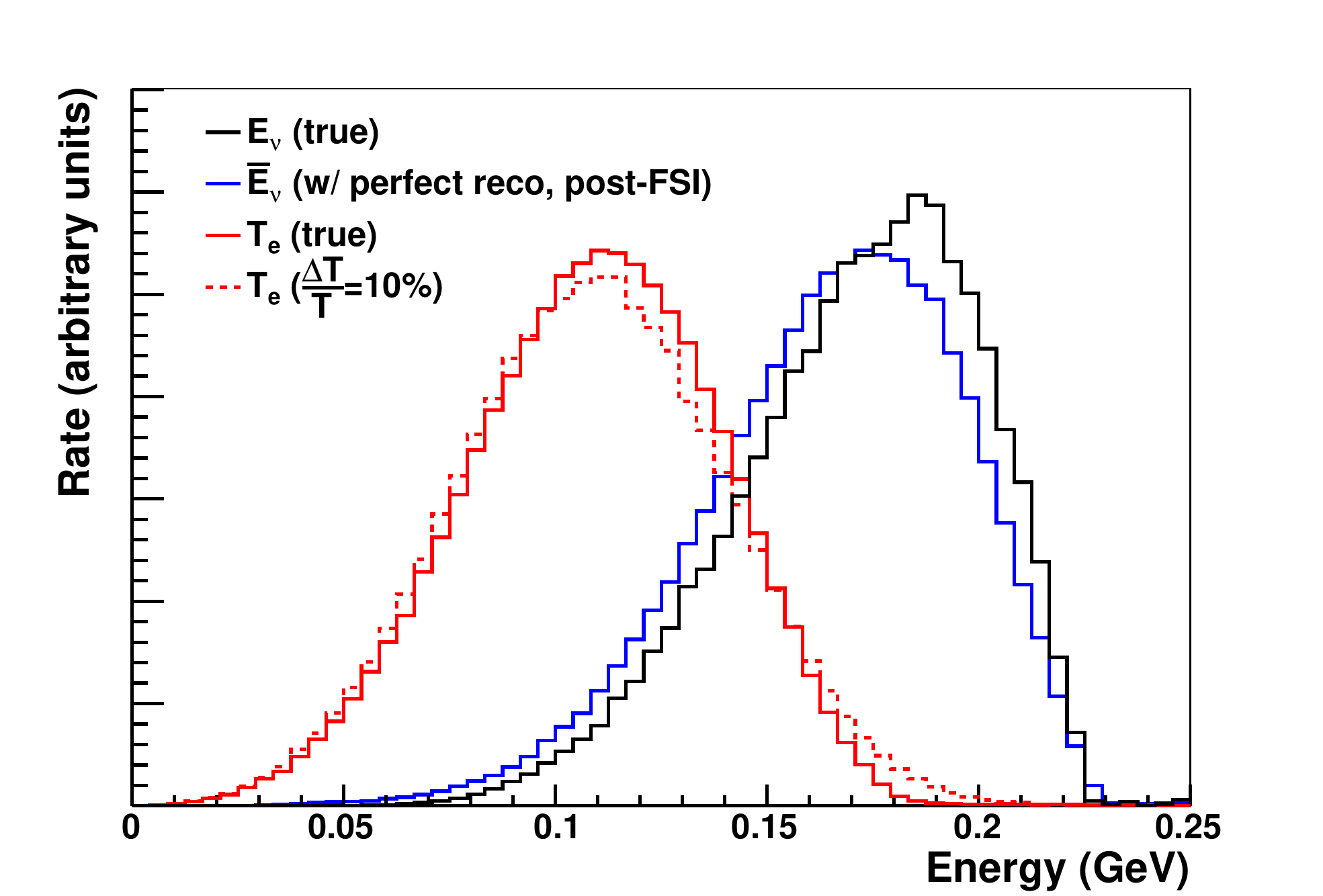} 
\caption{The NuWro simulation results of $\nu_e$ (from $K^{+} \rightarrow \pi^0  e^+  \nu_{e}$) CC interactions on carbon. The kinetic energy of the outgoing electron, given a set of detector resolution assumptions, is shown. The true neutrino energy and reconstructed neutrino energy available with a perfect detector ($\overline{E}_\nu$$=$$E_e + \sum_i^n T_{i,\mathrm{proton}} + S_p$) are also visible.}
\label{electron}
\end{centering}
\end{figure}

About 6500~$\nu_e$ CC events are expected from 100-225~MeV in true neutrino energy in four years of running the JPARC-MLF 50~ton LS experiment. This estimate uses the $\nu_e$ cross section prediction from Ref.~\cite{kolbe}. This sample may also be important as an experimental check of the $\nu_\mu$/$\nu_e$ cross section ratio, recalling that $\nu_\mu$ events are used to constrain the $\nu_e$ appearance expectation in both short- and long-baseline experiments. Notably, this ratio has been identified as one of the keys to improving sensitivity to $\delta_{\mathrm{CP}}$ in future long-baseline experiments~\cite{huberpaper} and our knowledge of it is weakest at low energy where the lepton mass difference and nuclear form factors contribute more. Measurements of these cross sections are also applicable in understanding the MiniBooNE low-energy excess at 250-475~MeV~\cite{mini_lowe} and for informing MicroBooNE's study of this energy region and below, given the low(er) energy reconstruction capabilities of LArTPC technology.  The results of a simulation of the $\nu_e$ from the JPARC-MLF source (shown in Fig.~\ref{mlf_flux}), which is basically the flux exclusively coming from $K^{+} \rightarrow \pi^0  e^+  \nu_{e}$, on a carbon target are shown in Fig.~\ref{electron}. The true neutrino energy, reconstructed neutrino energy available with a perfect detector $\overline{E}_\nu$ ($=$$E_e + \sum_i^n T_{i,\mathrm{proton}} + S_p$), and electron kinetic energy with two different detection resolution scenarios are shown. LS, LArTPC, and water Cerenkov technology are all capable of efficiently reconstructing $\nu_e$ CC events in this energy range.

It is also worth briefly mentioning that monoenergetic $\nu_\mu$-induced NC events may offer an interesting physics sample as well, especially since NC measurements are always flux-integrated; these may be useful for measuring $\Delta_{s}$ and/or isovector couplings in elastic scattering. 

\section{Conclusion}

Measurements of the monoenergetic 236~MeV $\nu_\mu$ from charged kaon decay-at-rest may be quite valuable for probing the nucleus using neutrinos and accelerator-based oscillation experiments. A number of experiments coming online in the next few years, including MicroBooNE at Fermilab and the LS-based experiment at JPARC's MLF facility, will be able to make precise cross section measurements of this unique known-energy channel. In particular, the MLF experiment will see close to 200,000~236~MeV $\nu_\mu$ CC events in four years of running along with 6500~$\nu_e$ CC events coming from a well-understood flux shape in the 100-225~MeV range. In the future, it seems pertinent to develop a quantitative understanding of the role these measurements can play in reducing the systematics associated with both energy reconstruction and cross sections in accelerator-based oscillation experiments, and perhaps even informing decisions related to the development of these programs.  

\section{Acknowledgements}
The author wishes to thank J.~Sobczyk and T.~Golan for help with NuWro; J.M.~Conrad for support and guidance; Z.~Pavlovic and NOvA for the NuMI flux simulation used in the MicroBooNE event rate estimate; T.~Maruyama and K. Nishikawa for providing a simplified geometry of the JPARC-MLF neutrino source that was used in the event rate estimate; T.~Katori for valuable discussions and comments; and G.P.~Zeller, F.~Cavanna, M.J.~Wilking, W.C.~Louis, B.J.P.~Jones, M. Moulai, and B.T.~Fleming for comments on early drafts. The author is supported by a Pappalardo Fellowship in Physics at MIT and by the National Science Foundation under Grant Number PHY-1205175.

\end{document}